\documentclass[iop,apj]{emulateapj}
\submitted{Accepted for publication in The Astrophysical Journal}
\begin{document}

\title{Transition to the disk dominant state of a new ultraluminous X-ray source in M82}

\author{Jing Jin\altaffilmark{1}, Hua Feng\altaffilmark{1}, \& Philip Kaaret\altaffilmark{2}}

\altaffiltext{1}{Department of Engineering Physics and Center for Astrophysics, Tsinghua University, Beijing 100084, China}
\altaffiltext{2}{Department of Physics and Astronomy, University of Iowa, Van Allen Hall, Iowa City, IA 52242, USA}

\shortauthors{Jin, Feng, \& Kaaret}
\shorttitle{A new ULX in M82}

\begin{abstract}
We report on the identification of a third, new ultraluminous X-ray source (ULX) in the starburst galaxy M82. Previously, the source was observed at fluxes consistent with the high state of Galactic black hole binaries (BHBs).  We observe fluxes up to $(6.5 \pm 0.3) \times 10^{39}$~ergs~s$^{-1}$ in the ultraluminous regime. When the source is not in the the low/hard state, spectral fitting using a multicolor disk model shows that the disk luminosity varies as the disk inner temperature raised to the power $4.8 \pm 0.9$, consistent with the behavior of Galactic BHBs in the thermal dominant state. Fitting the spectrum with a multicolor disk model with general relativistic corrections suggests that the source harbors a rapidly spinning black hole with a mass less than 100 solar masses. A soft excess was found in the source spectrum that could be blackbody emission from a photosphere created by a massive outflow. The source also showed soft dips during a flare. 
\end{abstract}


\section{Introduction}

Ultraluminous X-ray sources (ULXs) are nonnuclear, point-like X-ray sources in external galaxies with an apparent luminosity over the Eddington limit of a stellar-mass black hole, which is typically $3 \times 10^{39}$ ergs~s$^{-1}$ for a 20 $M_\sun$ black hole. Those with variability on time scales from minutes to months are accreting compact objects. The nature of ULXs is still unclear. If the emission is (roughly) isotropic and under the Eddington limit, they may harbor black holes up to $10^3$~$M_\sun$, belonging to the missing population of intermediate-mass black holes \citep[IMBHs;][]{col99,mak00,kaa01,far09}. However, they could also be stellar-mass black holes, with beamed or super-Eddington emission \citep{kin01,kor02,wat01,beg02}. 

ULXs with relatively low luminosities ($\sim 3-6 \times 10^{39}$~ergs~s$^{-1}$) may be more likely to be stellar-mass black holes of about 10~$M_\sun$ like Galactic BHBs or up to a couple tens of $M_\sun$ like IC 10 X-1 \citep{pre07,sil08}. Galactic BHBs have been observed in a few cases with luminosities up to $6 \times 10^{39}$ ergs~s$^{-1}$ \citep{mcc06}, although there are large uncertainties on their distances. \citet{gla09} proposed that most ULXs are stellar-mass black holes in an `ultraluminous state' with super-Eddington accretion based on analysis of high quality X-ray spectra. A major difference between the two populations is that Galactic BHBs spend most of their lifetime in the quiescent or low state, while most ULXs have persistent emission and have been active since their discovery at time scales up to decades in some cases. The non-transient behavior of the majority of ULXs implies that they do not contain black holes significantly more massive than $10^2$~$M_\sun$ \citep{kal04}. Despite these hypotheses, the connection between ULXs and Galactic BHBs has not been well addressed. If they contain black holes of similar mass, we may expect to see transitions from the normal states to the ultraluminous state in an individual source.
 
Multicolor blackbody emission from an optically thick accretion disk could shed light on the nature of the central black hole. At a fixed fraction of Eddington luminosity, the disk inner temperature scales with the black hole mass to the $-1/4$ power \citep{mak00}, indicating that accretion disks around IMBHs are cooler and brighter than those around stellar-mass black holes. Soft excesses are detected in ULXs and can be modeled as cool disk emission from IMBHs \citep{kaa03}. However, the cool disk model is not the unique interpretation of the soft excess \citep[for a brief review see the introduction of][]{fen09}. Massive outflows due to near or super-Eddington accretion may produce blackbody emission at the similar temperature \citep{kin03,kin04,beg06,pou07}.

The disk model can be tested by determining if the disk luminosity varies with the 4th power of the inner disk temperature, i.e.\ $L_{\rm disk} \propto T_{\rm in}^4$.  This relation has been robustly verified for stellar-mass black holes when the disk emission dominates the X-ray spectrum \citep{gie04}. Although the X-ray spectra of some ULXs can be modeled by dominant disk emission \citep{sto06,win06}, the observed luminosity variation and number of observations have been inadequate for proper testing of the $L_{\rm disk} \propto T_{\rm in}^4$ relation. A possible $L \propto T_{\rm in}^4$ relation was found in NGC 5204 X-1, but the disk emission was not the dominant component in the spectrum and a large correction for Comptonization was required \citep{fen09}. 

M82 is a nearby starburst galaxy in which two ULXs have been found \citep{mat01,kaa01,fen07b,kon07}. It has been extensively studied in X-rays with the Chandra X-ray observatory . In our recent Chandra observation on 2008 October 4, the X-ray source CXOM82 J095546.6+694041 was 5-10 times brighter than its normal flux level and was identified as the third ULX in M82. According to the naming convention suggested by \citet{kaa01}, we refer to this source as X37.8+54. In this paper, we investigated its spectral and timing behavior using all available Chandra observations, and discussed its possible nature. The distance to the host galaxy is adopted as 3.63 Mpc \citep{fre94}.

\section{Observations and Data Analysis}

So far there have been fourteen observations of M82 with the Chandra Advanced CCD Imaging Spectrometer (ACIS), including the most recent three that we proposed to monitor the brightest ULX in it (the results will be reported separately). Among them, eleven observations with an exposure longer than 5~ks are used for spectral analysis (Table~\ref{tab:obs}). The other three short ACIS observations (Obs ID 378, 380-1, and 380-2; 380 has two observations) as well as the four High Resolution Camera (HRC) observations (Obs ID 1411-1, 1411-2, 8505 and 8189; 1411 has two observations) are examined for long term variability. All data are processed using CIAO 4.1.2 with CALDB 4.1.3; new bad pixel files and level 2 events files are recreated when necessary.

\begin{deluxetable}{cccc}
\tablewidth{\columnwidth}
\tablecaption{Chandra ACIS Observations of M82 used for spectral analysis in the paper.
\label{tab:obs}}
\tablehead{
\colhead{ObsID} & \colhead{Date} & \colhead{Instrument} & \colhead{Exposure}
}
\startdata
361 & 1999-09-20 & ACIS-I & 33.3 \\
1302 & 1999-09-20 & ACIS-I & 15.5 \\
379 & 2000-03-11 & ACIS-I & 8.9 \\
2933 & 2002-06-18 & ACIS-S & 18.0 \\
6097 & 2005-02-04 & ACIS-S & 52.8 \\
5644 & 2005-08-17 & ACIS-S & 68.1 \\
6361 & 2005-08-18 & ACIS-S & 17.5 \\
8190 & 2007-06-02 & ACIS-S  & 52.8 \\
10027 & 2008-10-04 & ACIS-S & 18.3 \\
10025 & 2009-04-17 & ACIS-S & 17.4 \\
10026 & 2009-04-29 & ACIS-S & 16.9
\enddata
\end{deluxetable}

\begin{figure}
\plotone{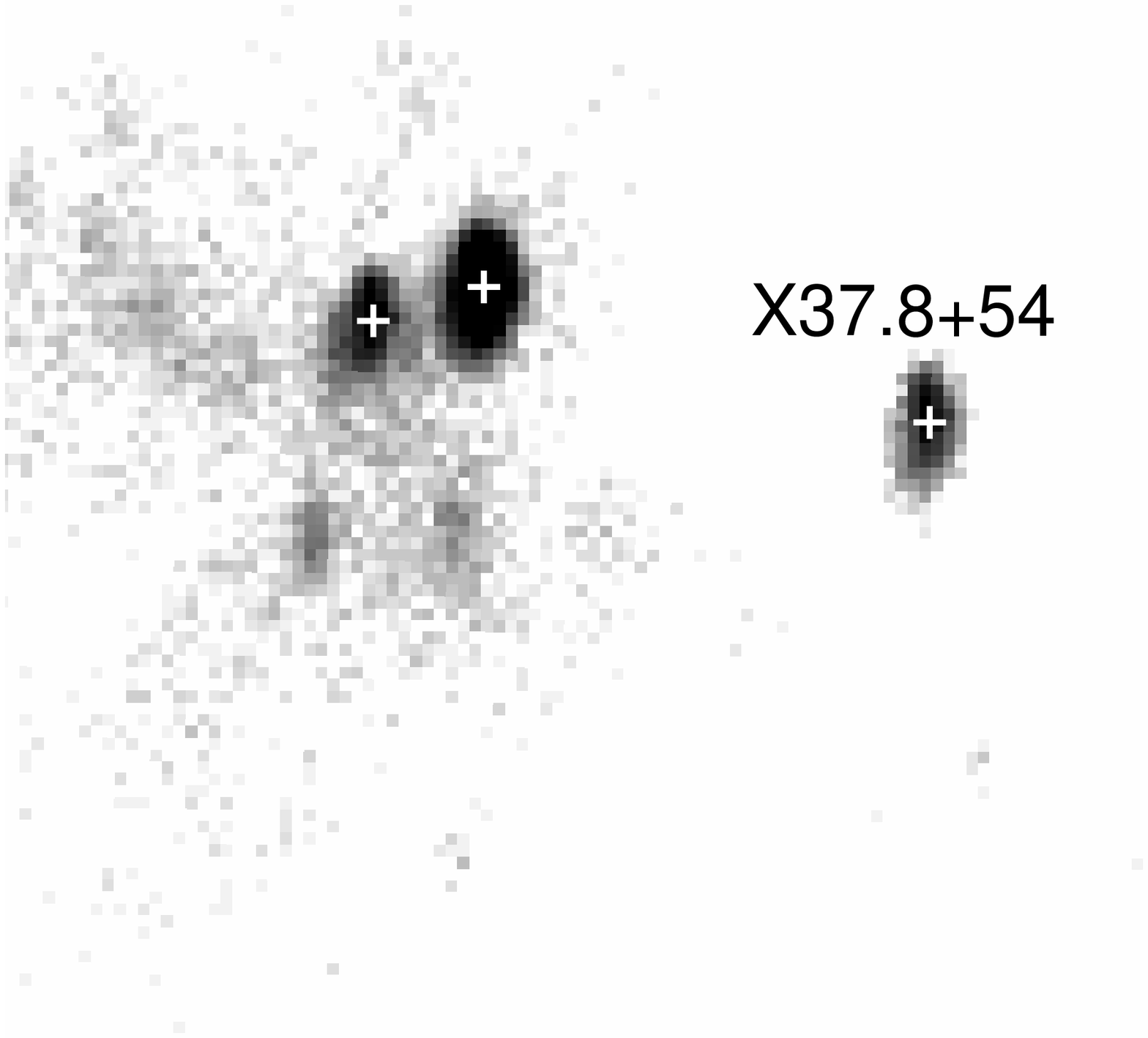}\\
\plotone{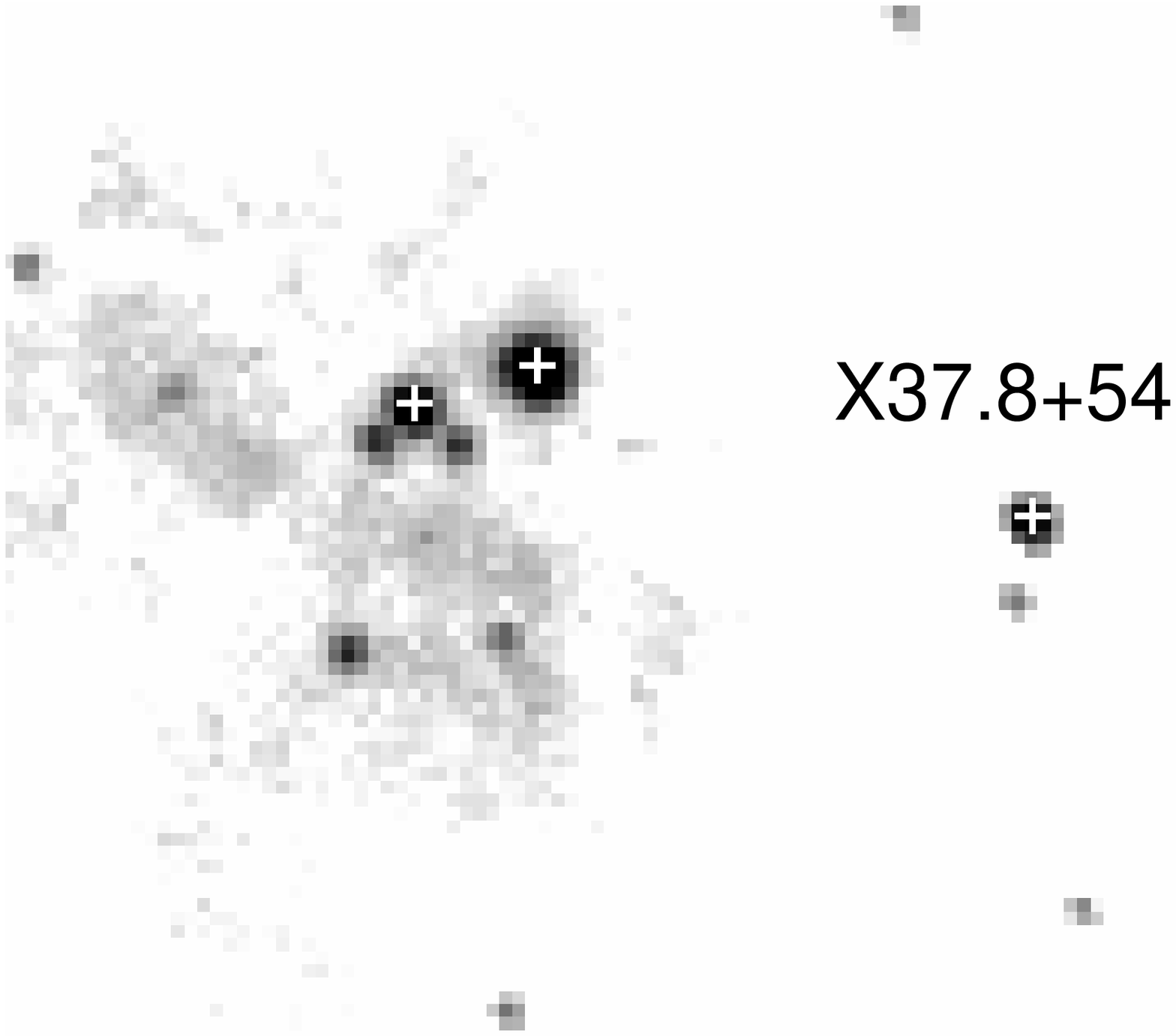}
\caption{
Chandra ACIS image of the central part of M82 in the energy range of 0.3-8 keV from observations 10027 ({\it top}) and 5644 ({\it bottom}), respectively, where X37.8+54 was in outbursts. The pluses indicate the three ULXs in M82. 
\label{fig:img}}
\end{figure}

Two images of the central region of M82 in the 0.3-8 keV energy range taken from Chandra ACIS are displayed in Figure~\ref{fig:img}. The top panel shows the image from observation 10027 when X37.8+54 was the brightest ever seen, and the bottom one is from observation 5644 in which the second brightest flux was recorded. The source in 10027 spreads on multiple pixels because it lies at a large off-axis angle. Aligning the two images using isolated point-like sources suggests that the brightening in observation 10027 around the source region is indeed from X37.8+54. A faint object to the south of X37.8+54 is seen in observation 5644, and is partly blended with X37.8+54 only in observation 10027. It displays a variability by a factor of 2. For observation 10027, we subtracted the contribution of this dim source assuming a spectrum from its highest state and found no significant change in the spectral fitting for X37.8+54, indicative of negligible contamination.

\begin{deluxetable*}{lllllll}
\tablecolumns{7}
\tablewidth{\textwidth}
\tablecaption{Spectral parameters of X37.8+54 derived from the power-law model subject to interstellar absorption.
\label{tab:pl}}
\tablehead{
\colhead{ObsID} & \colhead{$N_{\rm H}$} & \colhead{$\Gamma$} & \colhead{$N_{\rm PL}$} & \colhead{$f_{\rm X}$} & \colhead{$L_{\rm X}$} &  \colhead{$\chi^2$/dof}\\
\colhead{(1)} & \colhead{(2)} & \colhead{(3)} & \colhead{(4)} & \colhead{(5)} & \colhead{(6)} & \colhead{(7)}
}
\startdata
361+1302 & $7.7_{-0.7}^{+0.7}$ & $2.60_{-0.10}^{+0.11}$ & $2.3_{-0.3}^{+0.3}$ & $0.302_{-0.011}^{+0.011}$ & $1.7_{-0.2}^{+0.3}$ & 60.9/54 \\
379 & $3.6_{-1.7}^{+1.4}$ & $2.0_{-0.3}^{+0.2}$ & $1.8_{-0.5}^{+0.5}$ & $0.57_{-0.06}^{+0.07}$ & $1.5_{-0.2}^{+0.3}$ & 6.4/12 \\
2933 & $3.2_{-0.6}^{+0.7}$ & $1.62_{-0.11}^{+0.12}$ & $0.92_{-0.13}^{+0.15}$ & $0.45_{-0.03}^{+0.03}$ & $0.97_{-0.06}^{+0.06}$ & 19.9/28 \\

6097 & $6.7_{-0.4}^{+0.4}$ & $2.99_{-0.11}^{+0.11}$ & $2.0_{-0.2}^{+0.2}$ & $0.200_{-0.008}^{+0.008}$ & $1.6_{-0.2}^{+0.3}$ & 37.9/50 \\
5644 & $7.0_{-0.2}^{+0.2}$ & $2.64_{-0.05}^{+0.05}$ & $5.9_{-0.3}^{+0.3}$ & $0.791_{-0.014}^{+0.014}$ & $4.4_{-0.2}^{+0.2}$ & 161.2/149 \\
6361 & $6.4_{-0.7}^{+0.8}$ & $2.68_{-0.16}^{+0.18}$ & $2.9_{-0.4}^{+0.5}$ & $0.39_{-0.02}^{+0.02}$ & $2.2_{-0.4}^{+0.5}$ & 20.3/25 \\
8190 & $4.9_{-0.3}^{+0.3}$ & $2.22_{-0.07}^{+0.07}$ & $1.75_{-0.12}^{+0.13}$ & $0.401_{-0.012}^{+0.012}$ & $1.35_{-0.07}^{+0.08}$ & 92.6/82 \\
10027 & $7.5_{-0.4}^{+0.4}$ & $1.88_{-0.06}^{+0.06}$ & $7.4_{-0.5}^{+0.6}$ & $2.22_{-0.05}^{+0.05}$ & $6.5_{-0.3}^{+0.3}$ & 122.4/115 \\
10025 & $2.6_{-0.6}^{+0.7}$ & $1.73_{-0.12}^{+0.14}$ & $0.98_{-0.14}^{+0.17}$ & $0.43_{-0.03}^{+0.03}$ & $0.95_{-0.06}^{+0.07}$ & 33.3/26 \\
10026 & $4.4_{-0.7}^{+0.8}$ & $1.77_{-0.13}^{+0.14}$ & $1.07_{-0.16}^{+0.19}$ & $0.41_{-0.03}^{+0.03}$ & $1.00_{-0.07}^{+0.09}$ & 19.2/23
\enddata
\tablecomments{
Col.~(1): Observation ID. 
Col.~(2): Absorption column density in units of ${10^{21}}$~cm$^{-2}$. 
Col.~(3): The power-law photon index.
Col.~(4): The power-law normalization at 1~keV in units of $10^{-4}$~photons~cm$^{-2}$~s$^{-1}$.
Col.~(5): Observed flux in 0.3-8.0 keV in units of $10^{-12}$~ergs~cm$^{-2}$~s$^{-1}$.
Col.~(6): Unabsorbed luminosity in 0.3-8.0 keV in units of $10^{39}$~ergs~s$^{-1}$ assuming istropic emission at a distance of 3.63~Mpc.
Col.~(7): Best-fit $\chi^2$ and degrees of freedom.}
\end{deluxetable*}

\begin{deluxetable*}{lllllllll}
\tablecolumns{9}
\tablewidth{\textwidth}
\tablecaption{Spectral parameters of X37.8+54 fitted with the MCD model or the MCD plus blackbody model subject to interstellar absorption.
\label{tab:mcd}}
\tablehead{
\colhead{ObsID} & \colhead{$N_{\rm H}$} & \colhead{$T_{\rm in}$} & \colhead{$R_{\rm in}\sqrt{\cos i}$} & \colhead{$T_{\rm bb}$} & \colhead{$f_{\rm X}$} & \colhead{$L_{\rm disk}\cos i$} & \colhead{$L_{\rm bb}$} &  \colhead{$\chi^2$/dof}\\
\colhead{(1)} & \colhead{(2)} & \colhead{(3)} & \colhead{(4)} & \colhead{(5)} & \colhead{(6)} & \colhead{(7)} & \colhead{(8)} & \colhead{(9)}
}
\startdata
\multicolumn{8}{c}{Model: MCD}\\ \noalign{\smallskip}\hline\noalign{\smallskip}
379 & $0.7_{-0.2}^{+1.1}$ & $1.22_{-0.17}^{+0.17}$ & $4.1_{-0.8}^{+1.5}$ & \nodata & $0.50_{-0.06}^{+0.05}$ & $0.5_{-0.3}^{+0.5}$ & \nodata & 6.8/12 \\
2933 & $1.1_{-0.5}^{+0.4}$ & $1.74_{-0.15}^{+0.28}$ & $1.8_{-0.4}^{+0.3}$ & \nodata & $0.42_{-0.03}^{+0.03}$ & $0.4_{-0.2}^{+0.3}$ & \nodata & 27.0/28 \\
{\bf 10027} & $4.6_{-0.2}^{+0.3}$ & $1.51_{-0.06}^{+0.06}$ & $5.9_{-0.4}^{+0.5}$ & \nodata & $2.04_{-0.05}^{+0.05}$ & $2.3_{-0.5}^{+0.5}$ & \nodata & 123.5/115 \\
10025 & $0.5_{-0}^{+0.4}$ & $1.54_{-0.16}^{+0.13}$ & $2.3_{-0.3}^{+0.5}$ & \nodata & $0.40_{-0.03}^{+0.02}$ & $0.37_{-0.17}^{+0.21}$ & \nodata & 37.4/26 \\
10026 & $2.2_{-0.6}^{+0.4}$ & $1.48_{-0.12}^{+0.22}$ & $2.5_{-0.6}^{+0.4}$ & \nodata & $0.37_{-0.03}^{+0.03}$ & $0.37_{-0.19}^{+0.27}$ & \nodata & 23.8/23 \\
\cutinhead{Model: MCD + blackbody}
{\bf 361+1302} & $12._{-2.}^{+3.}$ & $1.02_{-0.08}^{+0.10}$ & $5.2_{-1.1}^{+1.2}$ & $0.163_{-0.019}^{+0.022}$ & $0.291_{-0.012}^{+0.007}$ & $0.38_{-0.18}^{+0.24}$ & $2.8_{-1.9}^{+4.6}$ & 51.6/52 \\
{\bf 6097} & $5.3_{-0.6}^{+0.7}$ & $0.88_{-0.10}^{+0.15}$ & $5._{-2.}^{+2.}$ & $0.21_{-0.05}^{+0.03}$ & $0.186_{-0.007}^{+0.004}$ & $0.21_{-0.16}^{+0.23}$ & $0.33_{-0.14}^{+0.46}$ & 36.5/48 \\
{\bf 5644} & $4.5_{-0.3}^{+0.4}$ & $1.185_{-0.013}^{+0.014}$ & $5.32_{-0.42}^{+0.08}$ & $0.290_{-0.027}^{+0.012}$ & $0.764_{-0.016}^{+0.009}$ & $0.72_{-0.11}^{+0.04}$ & $0.60_{-0.07}^{+0.08}$ & 151.4/147 \\
{\bf 6361} & $7.0_{-1.1}^{+2.3}$ & $1.11_{-0.16}^{+0.27}$ & $4.4_{-1.7}^{+1.8}$ & $0.20_{-0.04}^{+0.06}$ & $0.378_{-0.021}^{+0.013}$ & $0.4_{-0.3}^{+0.5}$ & $1.0_{-0.6}^{+1.8}$ & 17.5/23 \\
8190 & $4.9_{-0.7}^{+0.7}$ & $1.59_{-0.14}^{+0.15}$ & $2.1_{-0.4}^{+0.4}$ & $0.225_{-0.018}^{+0.021}$ & $0.395_{-0.008}^{+0.012}$ & $0.36_{-0.17}^{+0.21}$ & $0.48_{-0.11}^{+0.20}$ & 80.8/80 
\enddata
\tablecomments{
Col.~(1): Observation ID; those in bold font are not in the low/hard state. 
Col.~(2): Absorption column density in units of $10^{21}$~cm$^{-2}$. 
Col.~(3): Disk inner temperature in units of keV.
Col.~(4): $R_{\rm in}$ is the disk inner radius in units of 10~km and $i$ is the disk inclination angle.
Col.~(5): Blackbody temperature in units of keV.
Col.~(6): Observed flux in 0.3-8.0 keV in units of $10^{-12}$~ergs~cm$^{-2}$~s$^{-1}$.
Col.~(7): Disk bolometric luminosity in units of $10^{39}$~ergs~s$^{-1}$.
Col.~(8): Bolometric luminosity of the blackbody component in units of $10^{39}$~ergs~s$^{-1}$.
Col.~(9): Best-fit $\chi^2$ and degrees of freedom.}
\end{deluxetable*}


The energy spectra of X37.8+54 are extracted from events in a 3 $\sigma$ elliptical region around the source found by {\tt wavdetect}. The background is subtracted from a nearby circular region with a radius of about 4 arcsec on the same CCD and off the readout column of the source. Background lightcurves are checked using 10-15 keV events on the whole CCD; intervals with background flares were excluded in observations 5644 and 6361. Spectral fitting was done using XSPEC 12.5.0 and the errors are quoted at 68\% confidence level if not specified otherwise. Observations 361 and 1302 were performed in the same day and the spectral shapes are consistent. These two spectra are fitted together to improve the signal to noise ratio. The Galactic absorption column density along the line of sight to the source is $5.05 \times 10^{20}$~cm$^{-2}$ \citep{kal05}, and is set as the lower boundary of the total absorption column density during the fits. Using the {\tt pileup\_map} tool available in CIAO, we obtained that the count rate on the brightest 3 by 3 pixels is $\lesssim 0.1$ counts per frame. For the two observations 5644 and 10027 when the source was brightest, the highest count rate in a 3 by 3 pixel island is about 0.03 counts per frame due to a large off-axis angle and/or a small readout time. The derived spectral parameters do not change within errors when a pileup model is applied. Therefore, the effects of pileup are insignificant and, thus, not taken into account. 

\begin{figure}
\centering
\plotone{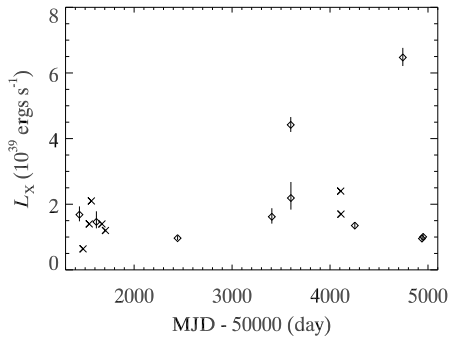}
\caption{
Long term X-ray lightcurve of X37.8+54 from all available Chandra observations. The diamond indicates the long ACIS observations and the luminosity is quoted from Table~\ref{tab:pl}. The cross indicates short ACIS and HRC observations and the luminosity is estimated from detected count rates using PIMMS.
\label{fig:longlc}}
\end{figure}

First, we tried a power-law model subject to interstellar absorption on each spectrum from observations listed in Table~\ref{tab:obs}. The best-fit parameters are listed in Table~\ref{tab:pl}. The source luminosity was found at two different levels, a relatively low level of about $(1-2) \times 10^{39}$~ergs~s$^{-1}$ and a high level above $4 \times 10^{39}$~ergs~s$^{-1}$. The long time scale lightcurve from all of the eighteen Chandra observations are shown in Figure~\ref{fig:longlc}. For the three short ACIS and the four HRC observations, which are inadequate for spectral fitting, their 0.3-8.0 keV luminosities are calculated using PIMMS from count rates, assuming a power-law spectrum with the average absorption column density and power-law photon index. They are found to be consistent with the relatively low flux level.

\begin{figure}
\centering
\plotone{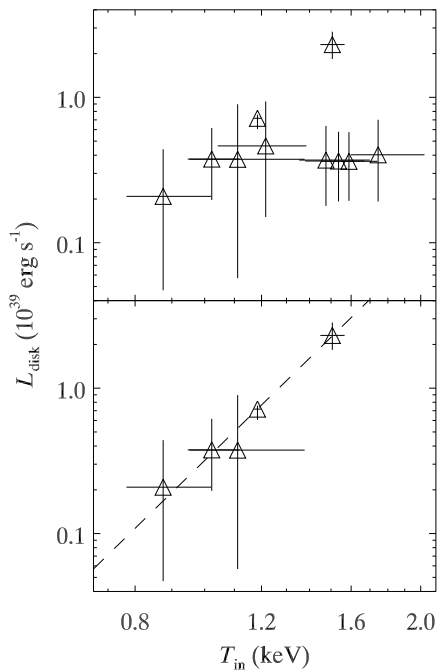}
\caption{
The disk bolometric luminosity (assuming face-on) versus the disk inner temperature of X37.8+54 with values obtained from the best-fit model in Table~\ref{tab:mcd}. In the bottom panel, data points from the five observations in which the source flux is low and the spectrum is hard are removed. The dashed line indicates a best-fit power-law relation $L_{\rm disk} \propto T_{\rm in}^{4.8 \pm 0.9}$ to the data.
\label{fig:mcd}}
\end{figure}

We then tried a multicolor accretion disk (MCD) model. A single MCD model with absorption did not improve the fits with respect to the power-law model. For five out of the ten spectra, an additional soft blackbody component to the MCD was justified at a significance of 3 $\sigma$ or more.  Spectral parameters derived from the MCD model or the MCD plus blackbody are listed in Table~\ref{tab:mcd}. The bi-thermal model produced comparable or better fits than the power-law model. The disk luminosity versus disk inner temperature is plotted in the top panel of Figure~\ref{fig:mcd}. The soft blackbody component mainly lies in the low energy band, and the diskbb component contributes more than 90\% of the unabsorbed flux in the 2.0-8.0 keV band. We note that four of the five brightest observations require the soft component, but the brightest observation, 10027, does not.  If we add a blackbody component into the 10027 spectrum and fix the temperature at its mean value of 0.2~keV, the blackbody luminosity is $0.5 \times 10^{39}$~ergs~s$^{-1}$, similar to those from other observations, and it contributes only 2.3\%, with a up limit of 5\%, to the total observed flux. The MCD parameters are consistent with those from a single MCD model. In the other four spectra, the fraction of the soft blackbody is larger than 18\%. The non-detection of the cool blackbody component in 10027 could be due to its relative weakness when the MCD component became much stronger than in others. 

\begin{deluxetable}{cc}
\tablecolumns{2}
\tablewidth{\columnwidth}
\tablecaption{Spectral parameters of X37.8+54 from observation 10027 fitted with the KERRBB model subject to interstellar absorption.
\label{tab:kb}}
\tablehead{\colhead{Parameters} & \colhead{Values}}
\startdata
$N_{\rm H}$ ($10^{21}$~cm$^{-2}$) & $5.0 \pm 0.3$ \\
$M_{\rm BH}$ ($M_\sun$) & $36 \pm 20$ \\
$\dot{m}$ ($10^{18}$ g s$^{-1}$) & $16.2_{-0.4}^{+91.0}$ \\
$a_\ast$ & $0.9986$ $(>0.80)$ \\
$i$ (degree) & $52 \pm 16$ \\
$\chi^2/\rm{dof}$ & 118.6/113 
\enddata
\tablecomments{
Torque at the inner edge of the disk is zero, the hardening correction is 1.7, the distance to the source is 3.63 Mpc, and the self-irradiation and limb-darkening are turned on.}
\end{deluxetable}

Except for observations in which the source spectra are hard and the luminosities are the lowest, the disk luminosity seems to be correlated with the disk inner temperature. In the bottom panel of Figure~\ref{fig:mcd}, we removed points from observations when the source was in the low/hard state, specifically when $L_{\rm X} < 2 \times 10^{39}$ ergs~s$^{-1}$ and $\Gamma < 2.3$, and then fitted the remaining observations to a power-law function. There appears to be a robust correlation and the best-fit relation between the luminosity and temperature is $L_{\rm disk} = T_{\rm in}^{4.8 \pm 0.9}$ with 1 $\sigma$ error. For the five observations (379, 2933, 8190, 10025, 10026) in which the spectrum is hard and the luminosity is relatively low, the source may be in the hard state with a power-law spectrum. Due to low statistics, these spectra can also be fitted by thermal disk radiation, but it is not surprising that they do not vary in a $L_{\rm disk} \propto T_{\rm in}^4$ pattern.

The MCD model has been widely used as a standard indicator of the thermal dominant state. However, it does not include relativistic effects. We thus tried a more physical multicolor disk blackbody model with relativistic effects fully taken into account \citep[KERRBB in XSPEC;][]{li05}, to the spectrum from observation 10027 when the source was brightest. In the model, we set zero torque at the inner boundary of the disk, a hardening correction of 1.7, a distance of 3.63 Mpc, and turned on self-irradiation and limb-darkening. The KERRBB model provides adequate fits to the spectrum with parameters (absorption column density, black hole mass, mass accretion rate, spin, and inclination) listed in Table~\ref{tab:kb}.

Models consisting of a power-law with an additional soft blackbody or cool MCD component were also fitted to the spectra. Neither showed obvious improvement relative to a single power-law model for most observations.

\begin{figure}
\centering
\plotone{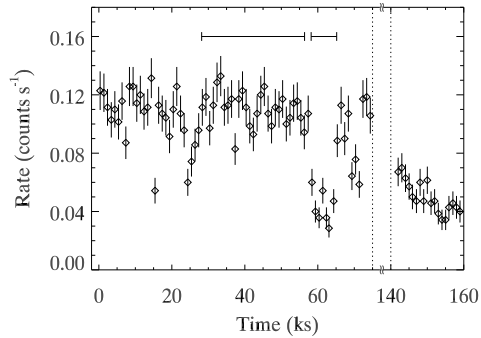}
\caption{
Lightcurve of X37.8+54 in 0.3-8 keV created from intervals when the
source is at least two pixels away from the readout node boundary.
The horizontal bars indicate the intervals used for
extracting energy spectra during the nondip and dip times, respectively. Simulations using MARX indicates that the variations seen in the figure are not from the instrument.  
\label{fig:lc}}
\end{figure}

\begin{figure}
\centering
\plotone{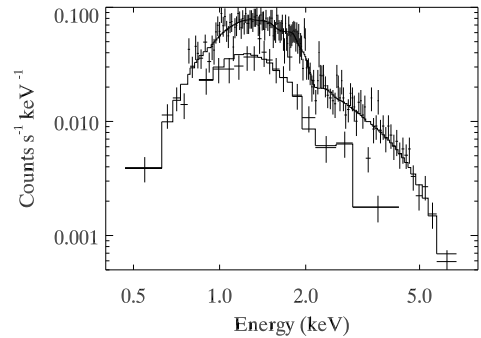}
\caption{
The dip (lower) and nondip (upper) spectra with their best fitted blackbody plus MCD model from observation 5644.
\label{fig:spe}}
\end{figure}

For each observation, the short-term variability was examined using the Kolmogorov-Smirnov test. Variability in the 0.3-8 keV range was found only in observations 5644 and 6361 with a significance of 9 and 5 $\sigma$, respectively. In all other observations, the evidence for variability is at a confidence level below 3 $\sigma$. We found that the source lies near the readout node boundary in observations 5644 and 6361, causing periodic oscillations when the source position on the detector plane moved across the node boundary due to dithering. The power spectrum shows a peak at 1000 s which is the dithering period along the direction perpendicular to the boundary. We therefore created a lightcurve from these two observations with time steps of about 1000~s selected from intervals when the source was at least two pixels away from the readout boundary. Interestingly, there are dip-like variations on time scales of several ks. Simulations using MARX 4.4.0 applying the aspect solution file and bad pixel map obtained from the observation indicates that dithering is not responsible for the variations shown in Figure~\ref{fig:lc}. We extracted energy spectra from intervals during the dip (58.171-65.171 ks) and the nondip (28.171-56.371 ks), which are marked in Figure~\ref{fig:lc}. We found that the dip spectrum is softer, with a power-law photon index of $3.5 \pm 0.4$ versus $2.65_{-0.09}^{+0.06}$ from the nondip spectrum. The spectrum during dip is also softer than in the low state, which has a power-law index less than 2.3. Fitting the nondip spectrum with a blackbody plus MCD model modified by a shared absorption component leads to consistent results with from the entire spectrum. We then use two absorption components applied onto each emission component individually. For the nondip spectrum, the absorption column density $N_{\rm H}$
on the MCD component is the same as on the blackbody component and is consistent with that derived using a single absorption component. For the dip spectrum, the fitting is made by fixing the temperature and $N_{\rm H}$ of the blackbody component to those of the nondip spectrum, due to limited photon number; other parameters, the normalization of the blackbody, the MCD and its absorption are free in the fit. This leads to a best-fit $N_{\rm H}$ of $40.3^{+17.5}_{-2.4}\times 10^{22}$~cm$^{-2}$ on the MCD component, which is significantly larger than the $N_{\rm H}$ from the nondip spectrum, and suggests that the dip may be caused by extra absorption of the MCD component. The dip and nondip spectra with this blackbody plus MCD model are shown in Figure~\ref{fig:spe} for comparison. Observation 6361 was performed 62 ks after the end of 5644. The source flux in 6361 seems to drop from the nondip level of 5644 to the relatively low level found in other observations.

We searched in the literature and found no infrared or radio counterpart for the X-ray source. The nearest radio source to X37.8+54 is 37.54+53.2 \citep{rod04} at an angular distance of $2''$ calculated using the X-ray position of R.A.=$9^{\rm h}55^{\rm m}46^{\rm s}.61$ and decl.=+$69^{\circ}40'41.1''$ (J2000.0) \citep{kaa01,fen07b}.

\section{Discussion}

We identify X37.8+54 as the third ULX in the starburst galaxy M82 based on its brightening in a recent Chandra observation to an observed flux 5-10 times higher than previously observed. The peak isotropic luminosity in the 0.3-8 keV band corrected for interstellar absorption is $(6.5 \pm 0.3) \times 10^{39}$~ergs~s$^{-1}$ inferred from fitting with a power-law model. 

The source was often found at relatively low luminosity, near $(1-2) \times 10^{39}$~ergs~s$^{-1}$. At the lowest luminosities, $\lesssim 1.5 \times 10^{39}$~ergs~s$^{-1}$, the source appears to have hard spectra with the photon index less than 2.3.  As the luminosity increases above this value, it shows dramatic spectral variation with the power-law photon index $\Gamma$ varying from 1.9 to 3.0.  This suggests a change in spectral state, and, indeed, most of the higher luminosity spectra are better fitted with an MCD spectral model.

The lower panel of Figure~\ref{fig:mcd} shows that when observations in the low/hard state are excluded, the remaining observations exhibit a robust correlation between $L_{\rm disk}$ and $T_{\rm in}$. The correlation coefficient between $\log L_{\rm disk}$ and $\log T_{\rm in}$ is 0.997 with a chance probability of $1.7\times 10^{-4}$. Fitting the data to a power-law relation, we obtained a power-law slope of $4.8 \pm 0.9$, consistent with the 4th-power relation predicted for the MCD model in the thermal dominant state. We note that the disk luminosity varies by about an order of magnitude.  Thus, the data provide an adequate test of this relation. We conclude that X37.8+54 exhibits two spectral states: a low/hard state and a thermal dominant state.
 
Following the recipe in \citet{mak00} and adopting canonical correction coefficients with $\kappa=1.7$ and $\xi=0.412$ ,where $\kappa$ is the hardening correction factor and $\xi$ is to correct for the radius of maximum temperature, the disk luminosity can be expressed as a function of the black hole mass ($m$ in solar mass) and the disk inner temperature ($T_{\rm in}$ in keV) as $L_{\rm disk} = 7.2\times 10^{36} \alpha^2 m^2 T_{\rm in}^4$~ergs~s$^{-1}$, where $\alpha$ is related to the black hole spin with $\alpha=1$ for a non-spinning black hole and $\alpha=1/6$ for maximal spinning. Using a 4th power-law relation to fit the data points in the bottom panel of Figure~\ref{fig:mcd}, we obtain $L_{\rm disk} = (3.7 \pm 0.3)T_{\rm in}^{4}/\cos i \; 10^{38}$~ergs~s$^{-1}$. Thus, we have $m = (7.2 \pm 0.3)/(\alpha\sqrt{\cos i})$, suggesting the ULX contains a black hole of about 10 solar masses if it is a Schwartzchild black hole with a face-on disk, or a few tens solar masses if it is fast spinning or viewed at a high inclination. 

The best-fit compact object mass is $36 \pm 20$~$M_\sun$, derived from fitting with the fully relativistic disk model, indicative of a stellar black hole slightly more massive than Galactic ones. We also tried different hardening correction factors from 1.5 to 1.9. The fitted mass varies by 20\% with respect to that from $\kappa=1.7$. Interestingly, the specific angular momentum converges to its maximally allowed value, $a_\ast = 0.9986$ with a lower limit of 0.80, suggesting the source contains a fast spinning black hole. The mass accretion rate in Eddington unit is about 1.0, which is larger than that found in the thermal dominant state of Galactic BHBs of no more than 0.6 \citep{mcc06}, but is under the theoretical upper limit of about 10 for a stellar-mass black hole with a radiation dominated accretion disk \citep{beg02}. \citet{hui08} investigated the spectra of six disk-dominated ULXs and found five sources had best-fit masses between 23 and 73 $M_\sun$. The three ULXs that have a derived mass $\gtrsim$ $25~M_\sun$ at a confidence $\gtrsim 99.9\%$ also show rapid spin, which is quite similar to X37.8+54. The best-fit black hole mass of X37.8+54 is also similar with that of IC~10~X-1 \citep{pre07,sil08}. It has been suggested that IC~10~X-1 has a persistently high accretion rate \citep{bar08}; a rapidly spinning black hole is naturally expected since accretion would bring a fair amount of angular momentum to the compact object.

We conclude that stellar Kerr black holes with a mass of several tens $M_\sun$ could appear as ULXs with a thermal spectrum of a few keV in the 1-10 keV band, as having been suggested by \citet{mak00}. Such emission could also appear as a hard power-law spectrum if the statistics is inadequate to detect the spectral curvature. These black holes are more massive than Galactic black holes, which are mostly around 10 $M_\sun$, and may be formed by the collapse of massive stars in a low-metallicity environment and account for a portion of ULXs \citep{zam09}.

Similar sources, Suzaku J1305-4931 in NGC 4945 and NGC 253 X-2, have been reported in the literature, which show a $L_{\rm disk} \propto T_{\rm in}^4$ correlation as their luminosities varied by a factor of about 2-3 \citep{iso08,kaj09}. From surveys of bright ULXs in nearby galaxies, many hard ULXs were found and interpreted as IMBHs in the low/hard state \citep{fen09,kaa09}. For those sources, most of them present a constant spectral index or hardness despite large variation of the luminosity. Therefore, those sources may have a different nature from X37.8+54. 

It has been suggested that ULXs are a special case (face-on) of the Galactic microquasar SS~433, where super-Eddington accretion drives a powerful wind off the accretion disk \citep{fab01,beg06}. The wind could be optically thick and produce a blackbody continuum \citep{kin03,kin04}. This may be responsible for the soft excess observed in many ULXs with a temperature of about 0.1-0.3 keV \citep{sto06,pou07,fen07a}. When fitting spectra not from the low/hard state with an MCD component, all observations except 10027 need an additional blackbody component at a significance of at least 3 $\sigma$. No obvious correlation was found between the luminosity and temperature of the soft blackbody.  However, the uncertainties are large and this result is not very constraining. The calculated surface area of the soft blackbody emission is about $10^{17}$~cm$^2$ for all observations, corresponding to a spheric radius of $10^8$~cm, about $70R_{\rm g}$ for a 10 solar mass black hole, where $R_{\rm g}$ is the gravitational radius equal to $GM/c^2$.  The absorption column density varies significantly between observations no matter what models are used. Such large variation is likely internal to the binary, implying the outflow is inhomogeneous and time variable.  We note that the soft blackbody appears at energies where the absorption is important. Thus caution is warranted when interpreting it.

Interesting timing behavior is seen in one of the major outbursts caught by the observations 5644 and 6361 (Figure~\ref{fig:lc}). The source shows soft dips on time scales of several ks. Soft dips could be caused by absorption of the inner accretion disk, like in 4U~1630$-$47 \citep{tom98}. This interpretation is also consistent with our spectral analysis results in the penultimate paragraph of Section 2. We note that the soft blackbody component is assumed not to suffer extra absorption. A dense massive outflow from the disk could be the origin of the extra absorption and also the source of the soft blackbody. Other mechanisms could also produce soft dips, e.g., the possible ejection of the inner part of accretion disk into relativistic jets in GRS~1915$+$105 \citep{nai01}. The time scale of ks of the dip in X37.8+54 is longer than that in 4U~1630$-$47 and in GRS~1915$+$105 by about an order of magnitude. More detailed study of the soft dip requires data with better quality.

To summarize, X37.8+54 provides us a good candidate of a stellar Kerr black hole being ultraluminous with a near Eddington accretion rate. This ULX could be representative of a class of ULXs with relatively low luminosity, i.e.\ a few $10^{39}$~ergs~s$^{-1}$. The luminosity of X37.8+54 is usually not high enough to be classified as a ULX, while in the outburst it truly enters the ultralumious regime. The X-ray spectrum at peak luminosity is dominated by a disk component, which is unlike many brighter ULXs at their maximum luminosity that they enter the `ultraluminous state' characterized by a strong Comptonizing component \citep{gla09}. Therefore, this source may be in between the low luminosity tail of ULXs and high luminosity tail of Galactic BHBs. We note that the hard state of X37.8+54 is more luminous than the canonical hard state found in Galactic BHBs. This is consistent with the correlation found in outbursts of BHBs that the peak luminosity of the hard state is scaled with the peak luminosity of the thermal state in the following outburst \citep{yu09}.

\acknowledgments We thank the referee for useful comments. HF acknowledges funding support from the National Natural Science Foundation of China under grant No.\ 10903004 and 10978001, the 973 Program of China under grant 2009CB824800, and the Foundation for the Author of National Excellent Doctoral Dissertation of China under grant 200935. PK acknowledges support from Chandra grant GO9-0034X.

\end{document}